\documentclass[a4paper,12pt]{article}

\usepackage{amsmath}
\usepackage[psamsfonts]{amssymb}
\usepackage{rsfs}
\usepackage{bm}

\usepackage{cite}
\usepackage[dvips]{graphicx}

\begin{document} 

\begin{titlepage}

\baselineskip 10pt
\hrule 
\vskip 5pt
\leftline{}
\leftline{Chiba Univ. Preprint
          \hfill   \small \hbox{\bf CHIBA-EP-137}}
\leftline{\hfill   \small \hbox{hep-th/0209237}}
\leftline{\hfill   \small \hbox{September 2002}}
\vskip 5pt
\baselineskip 14pt
\hrule 
\vskip 1.0cm
\centerline{\Large\bf 
Consistent power corrections 
} 
\vskip 0.3cm
\centerline{\Large\bf  
to ultraviolet asymptotic solutions 
}
\vskip 0.3cm
\centerline{\Large\bf  
in Yang-Mills theory
}
\vskip 0.3cm
\centerline{\large\bf  
}

\vskip 0.5cm

\centerline{{\bf 
Kei-Ichi Kondo$^{\dagger,{1}}$ 
}}  
\vskip 0.5cm
\centerline{\it
${}^{\dagger}$Department of Physics, Faculty of Science, 
Chiba University, Chiba 263-8522, Japan
}
\vskip 1cm

\begin{abstract}
We show that the $1/p^2$ power corrections to the ultraviolet asymptotic solutions are allowed  as consistent solutions of the coupled Schwinger-Dyson equation for the gluon and (Faddeev-Popov) ghost propagators in Yang-Mills theory.
This result supports the existence of the vacuum condensate $\langle A_\mu^2 \rangle$ with mass dimension 2, as recently suggested by the operator product expansion and lattice simulations.  
We compare the solution with the result of operator product expansion.

\end{abstract}


\vskip 0.2cm
\hrule  
\vskip 0.2cm
${}^1$ 
  E-mail:  {\tt kondo@cuphd.nd.chiba-u.ac.jp}

\vskip 0.2cm  

\par 
\par\noindent


\vskip 0.5cm

\pagenumbering{roman}

\vskip 0.5cm  



\end{titlepage}


\pagenumbering{arabic}

\baselineskip 14pt
\section{Introduction}

The vacuum condensate is known to play the distinguished role for characterizing the non-perturbative aspect of the quantum chromodynamics (QCD).  
In fact, the gauge invariant vacuum condensates such as gluon condensate $\langle \mathscr{F}_{\mu\nu}^2 \rangle$ of mass dimension 4 and chiral condensate $\langle \bar{\psi}\psi \rangle$ of mass dimension 3 are well-known examples studied extensively so far.   
  
Recently,  several groups \cite{GSZ01,Boucaudetal00,Kondo01,Boucaudetal02,VKAV01,Schaden99} have focused their attention to a novel type of vacuum condensate.
In a series of lattice simulations \cite{Boucaudetal00}, the gluon propagator and the running coupling constant of Yang-Mills theory (QCD without quarks) in the Landau gauge  has been computed at large momenta, and it was shown that these data are compatible with the existence of the rather large $O(1/p^2)$ correction to the perturbative predictions.
In an operator product expansion (OPE) approach \cite{Boucaudetal00,Kondo01}, this correction is related to the existence of $\mathscr{A}_\mu^2$ condensate in the Landau gauge.%
\footnote{
In the previous work \cite{Kondo01}, the OPE was performed among others in the most general manifestly Lorentz covariant gauge which  reduces to the Landau gauge in a special limit $\lambda \rightarrow 0$ in the notation of \cite{Kondo01}.  Therefore, the OPE result for the gluon propagator in the Landau gauge \cite{Boucaudetal00} was confirmed by \cite{Kondo01}, while the OPE of the Faddeev-Popov ghost propagator was calculated anew in \cite{Kondo01}.
}   
It was also claimed \cite{Boucaudetal02} that this condensate might be related to instantons.  
The numerical investigations \cite{Boucaudetal00,Boucaudetal02} reached the sizable value for the vacuum condensate $\langle \mathscr{A}_\mu^2 \rangle \cong (1.4 \text{GeV})^2$.

In the Maximal Abelian (MA) gauge \cite{tHooft81}, on the other hand, other types of vacuum condensate of mass dimension 2, i.e., off-diagonal gluon condensate $A_\mu^a A_\mu^a$ and off-diagonal ghost-antighost condensate $\bar{C}^aC^a$, have been proposed and investigated in relation to quark confinement \cite{Schaden99}.  These condensates can provide the non-zero mass for the off-diagonal gluons and ghosts, giving a natural explanation for the Abelian dominance confirmed by numerical simulations and quark confinement \cite{KondoI}.  

Although the composite operator $\mathscr{A}_\mu^2$ is not a gauge invariant quantity, it was pointed out that its minimum along the gauge orbit can have a definite meaning \cite{GSZ01}.
Moreover,  it was shown \cite{Kondo01} that $\mathscr{A}_\mu^2$ is regarded as a Landau gauge limit of the more general composite operator of mass dimension 2  which is invariant under the Becchi-Rouet-Stora-Tyutin (BRST) transformation  in the most general manifestly Lorentz covariant gauge.  
That is to say, the vacuum condensate of mass dimension 2 can be allowed to exist as a BRST invariant quantity.

The main purpose of the present paper is to give another evidence of the existence of $O(1/p^2)$ correction to the perturbative expressions which are expected to hold in the large momenta.
In the present paper, we study the existence of such a power correction in the ultraviolet (UV) asymptotic region within the framework of the truncated Schwinger-Dyson (SD) equation for the gluon and ghost propagators in Yang-Mills theory.  
In the previous paper  \cite{Kondo02}, we have examined whether  the coupled SD equation allows logarithmic corrections to the infrared (IR) asymptotic solutions with power momentum dependence  predicted by Gribov and Zwanziger \cite{Gribov78} and confirmed based on various methods; SD equation \cite{SHA97,AB98,AB98b,Bloch01,FAR02,LS02}, stochastic equation \cite{Zwanziger01}, lattice simulations \cite{Bonnetetal00}. 
Such logarithmic corrections are proved to be absent in IR asymptotic solution to the truncated SD equation \cite{Kondo02}, in sharp contrast to the UV asymptotic solution.
For the relationship between color confinement and the solution of the SD equation, see e.g., \cite{Kondo02}.
\par
In order to show that both the gluon and ghost propagators can have the $O(1/p^2)$ power correction to the  UV  asymptotic solution with logarithmic momentum dependence,   
we adopt the Ansatz for the gluon form factor $F(p^2)$ and ghost form factor $G(p^2)$
defined by the gluon propagator in the Landau gauge and the ghost propagator with unbroken color symmetry in the $SU(N_c)$ Yang-Mills theory,
\begin{align}
  D_{\mu\nu}^{AB}(p) := \delta^{AB} P_{\mu\nu}^T(p)  F(p^2)/p^2   ,
\quad 
 G^{AB}(p) := \delta^{AB}  G(p^2)/p^2   ,
\end{align}
with $A,B=1, \cdots, N_c^2-1$ and the transverse projection operator,
$
  P_{\mu\nu}^T(p) := \delta_{\mu\nu} -{p_\mu p_\nu \over p^2} ,
$
\begin{subequations}
\begin{align}
  F(z) =& A z^{\gamma} \sum_{n=0}^{N} c_n z^{-n}   
+ \sum_{\ell=1}^{N'} e^{-\ell z/\omega} z^{\tilde{\gamma}_\ell} a_0^{(\ell)}  ,
  \\
  G(z) =& B z^{\delta} \sum_{n=0}^{N} d_n z^{-n} 
+ \sum_{\ell=1}^{N'} e^{-\ell z/\omega} z^{\tilde{\delta}_\ell} b_0^{(\ell)}  ,
\end{align}
\label{ansatz}
\end{subequations}
where%
\footnote{We can consider more general Ansatz which  is equivalent to the sum of the logarithmic and the power forms with possible logarithmic corrections:
$
  F(z) = A z^{\gamma} \sum_{n=0}^{N} c_n z^{-n} 
   +    \sum_{\ell=1}^{N} e^{-\ell z/\omega} z^{\tilde{\gamma}_\ell}  \sum_{n=0}^{N} K_n^{(\ell)} z^{-n}   ,
$
where $\tilde{\gamma}_\ell :=\gamma+\gamma_\ell$ and 
$K_n^{(\ell)}:=A a_n^{(\ell)} c_n$. 
The first correction factor in the above Ansatz corresponds to the higher-order logarithmic corrections in perturbation theory, while the second factor  to the power corrections.  
}
 $c_0=1=d_0$. 
Here we have introduced a variable \cite{Kondo91,Kondo01},
\begin{align}
  z :=  \omega \ln {p^2 \over \sigma} + z_\sigma ,
\end{align}
where $\omega$ is a real number  and 
$\sigma$ is the renormalization group invariant momentum scale defined by
$\sigma:=\mu^2 \exp [-2 \int_{g_0}^{g} {d\lambda \over \beta(\lambda)}]$ for the renormalization scale $\mu$ using the $\beta$ function: 
$\beta(g):= \mu {dg(\mu) \over d\mu}$.
The UV asymptotic limit $p^2 \rightarrow \infty$ corresponds to $z \rightarrow \infty$.  

\section{Power corrections to asymptotic solutions of the coupled SD equation}

\begin{figure}[htbp]
\begin{center}
\includegraphics{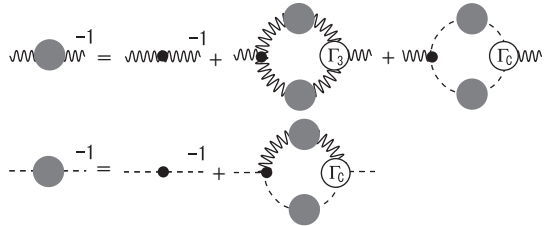}
\caption{Diagrammatic representation of the truncated Schwinger-Dyson equations for gluon and ghost propagators.}
\label{fig:SDeq}
\end{center}
\end{figure}

In this paper we deal with the renormalized version of the SD equation.  

0) We introduce the renormalization constants for the gluon field $Z_3$, the ghost field $\tilde{Z}_3$, the triple gluon vertex $Z_1$ and the gluon-ghost-antighost vertex $\tilde{Z}_1$ to write down the SD equation for the {\it renormalized} gluon and ghost form factors.  

First, we restrict our consideration to a version of the truncated SD equation 
subject to the following truncation, procedure and an approximation  
in order to write down the solvable SD equation. 
See Fig.~\ref{fig:SDeq} for the diagrammatic representation of the SD equation treated in the present paper.

1) 
The full vertex function is replaced by the bare one. 

2) 
Contributions from the two-loop diagrams are neglected in the gluon equation, while all diagrams are included in the ghost equation.  

3) 
The gluon equation is contracted with the Brown-Pennington projection operator \cite{BP88} 
$
  R_{\mu\nu}(p) := \delta_{\mu\nu} - 4{p_\mu p_\nu \over p^2} ,
$
to remove the quadratic UV divergence coming from the tadpole term.  

4) 
The $y$-max approximation:%
\footnote{This approximation was called Landau-Abrikosov-Khalatnikov (LAK) approximation or Higashijima-Miransky (HM) \cite{HM84} approximation.  
In QED, the angular integration in the coupled SD equation in the bare vertex approximation was first performed by Kondo, Mino and Nakatani \cite{KN92} and subsequently by Bloch and Pennington \cite{BP95} in a numerical way. 
In the pure Yang-Mills theory (QCD without quarks), the exact angular integration in the bare vertex approximation was performed in an analytical way for the pure power solution as the IR asymptotic solution \cite{AB98b} without relying on this approximation.
Recently,  a more sophisticated procedure was invented \cite{FAR02,LS02} for the solution to be
consistent with the renormalization effect of the triple gluon interaction vertex which is necessary to reproduce the correct coefficient of $\beta$ function in the UV limit. 
This issue will be discussed in the next section in more detail.  
}
$
 F((p-q)^2)=F(\text{max}\{p^2,q^2\}) , \quad
 G((p-q)^2)=G(\text{max}\{p^2,q^2\}) ,
$
is adopted to avoid angular integration of the angle $\theta$
where the angle $\theta$ comes from the inner product $p \cdot q= \sqrt{p^2} \sqrt{q^2} \cos \theta$ between the external momenta $p$ and the internal (loop) momenta $q$ which is to be integrated out.

By taking the approximations 1), 2) and 3), the SD equation includes only  two types of the bare vertex for the triple gluon and the gluon-ghost-antighost interactions.
The non-trivial contribution in the gluon equation comes from only two diagrams, i.e., gluon loop and ghost loop, apart from the trivial tree gluon propagator.  
In particular, the projection 3) removes the tadpole diagram.
On the other hand, the ghost equation includes a mixed loop composed of gluon and ghost.
By the approximation 4), the coupled SD equations are reduced to the one-dimensional integral equations with integration variable $q^2 \in [0,\Lambda^2]$ where $\Lambda$ is the UV cutoff.
Possible improvements of the above approximations are discussed in the next section.

Thus, the ghost SD equation reads
\begin{align}
  G^{-1}(x) = \tilde{Z}_3 - {3N_c \over 4}\lambda \tilde{Z}_1 \Big[ {F(x) \over x^2} \int_{0}^{x} dy y G(y) + \int_{x}^{\Lambda^2} {dy \over y} F(y) G(y) \Big] ,
\end{align}
while the gluon SD equation reads
\begin{align}
  F^{-1}(x) =& {Z}_3 + {N_c \over 3}\lambda \tilde{Z}_1 \Big[ -{G(x) \over x^3} \int_{0}^{x} dy y^2 G(y) + {3G(x) \over 2x^2} \int_{0}^{x} dy y G(y) 
  + {1 \over 2} \int_{x}^{\Lambda^2} {dy \over y} G^2(y)  \Big] 
\nonumber\\
&+ {N_c \over 3}\lambda {Z}_1 \Big[  {7F(x) \over 2x^3} \int_{0}^{x} dy y^2 F(y) 
- {17F(x) \over 2x^2} \int_{0}^{x} dy y F(y) 
- {9F(x) \over 8x} \int_{0}^{x} dy F(y) 
\nonumber\\
  &- 7 \int_{x}^{\Lambda^2} {dy \over y} F^2(y)
  + {7x \over 8} \int_{x}^{\Lambda^2} {dy \over y^2} F^2(y)  \Big] ,
\end{align}
where we have introduced 
$\lambda :={g^2 \over 16\pi^2}$, $x:=p^2$ and $y:=q^2$.

The two of the four renormalization constants, $Z_3$ and $\tilde{Z}_3$, can be eliminated by subtracting the equation at $x=\sigma$.
The ghost equation reads
\begin{align}
  G^{-1}(x) = G^{-1}(\sigma) - {3N_c \over 4}\lambda \tilde{Z}_1 \Big[ {F(x) \over x^2} \int_{0}^{x} dy y G(y) 
  - (x \rightarrow \sigma)
   + \int_{x}^{\sigma} {dy \over y} F(y) G(y) \Big] ,
\end{align}
while the gluon equation reads
\begin{align}
  F^{-1}(x) =& F^{-1}(\sigma) + {N_c \over 3}\lambda \tilde{Z}_1 \Big[ -{G(x) \over x^3} \int_{0}^{x} dy y^2 G(y) + {3G(x) \over 2x^2} \int_{0}^{x} dy y G(y) 
\nonumber\\
  & -(x \rightarrow \sigma)
  + \int_{x}^{\sigma} {dy \over 2y} G^2(y)  \Big] 
\nonumber\\
&+ {N_c \over 3}\lambda {Z}_1 \Big[  {7F(x) \over 2x^3} \int_{0}^{x} dy y^2 F(y) 
- {17F(x) \over 2x^2} \int_{0}^{x} dy y F(y) 
- {9F(x) \over 8x} \int_{0}^{x} dy F(y) 
\nonumber\\
& 
  + {7x \over 8} \int_{x}^{\Lambda^2} {dy \over y^2} F^2(y) 
  - (x \rightarrow \sigma)
  - 7 \int_{x}^{\sigma} {dy \over y} F^2(y) \Big]  .
\end{align}

The coupled SD equation obtained in this way agrees with the version adopted by 
Atkinson and Bloch \cite{AB98}.
It is known that this solution exhibits qualitatively the same IR behavior as the solutions of the truncated SD equations using the dressed vertex function improved so as to be consistent with the Slavnov-Taylor identity. 
Moreover, it has been shown that the leading logarithmic behavior is reproduced in the UV asymptotic solution so that it is consistent with the perturbation theory at one-loop order. In this paper, we consider the sub-leading contribution and power-corrections in the UV asymptotic solution.

\par
In the actual calculations, it is more convenient to rewrite the coupled SD equation in terms of the new variable 
$z:= \omega \ln {x \over \sigma} + z_\sigma$ 
and 
$\zeta:= \omega \ln {y \over \sigma} + z_\sigma$.  
Then the ghost equation reads
\begin{align}
  G^{-1}(z) - G^{-1}(z_\sigma) =& 
  - {3N_c \over 4}\lambda \tilde{Z}_1 \Big[  F(z) e^{-2z/\omega} \int_{-\omega \infty}^{z} {d\zeta \over \omega} e^{2\zeta/\omega} G(\zeta) 
\nonumber\\&
   - (x \rightarrow \sigma; z \rightarrow z_\sigma)
  + \int_{z}^{z_\sigma} {d\zeta \over \omega} F(\zeta) G(\zeta) \Big] ,
  \label{ghosteq}
\end{align}
and the gluon equation reads
\begin{align}
  & F^{-1}(z) - F^{-1}(z_\sigma) 
\nonumber\\ 
=& 
    {N_c \over 3}\lambda \tilde{Z}_1 \Big[ - G(z) e^{-3z/\omega} \int_{-\omega \infty}^{z} {d\zeta \over \omega} e^{3\zeta/\omega} G(\zeta) 
  + {3 \over 2} G(z) e^{-2z/\omega} \int_{-\omega \infty}^{z} {d\zeta \over \omega} e^{2\zeta/\omega} G(\zeta) 
\nonumber\\
  & -(x \rightarrow \sigma; z \rightarrow z_\sigma)
  + {1 \over 2} \int_{z}^{z_\sigma} {d\zeta \over \omega}  G^2(\zeta)  \Big] 
\nonumber\\
&+ {N_c \over 3}\lambda {Z}_1 \Big[  {7 \over 2} F(z) e^{-3z/\omega} \int_{-\omega \infty}^{z} {d\zeta \over \omega} e^{3\zeta/\omega} F(\zeta) 
- {17 \over 2} F(z) e^{-2z/\omega} \int_{-\omega \infty}^{z} {d\zeta \over \omega} e^{2\zeta/\omega} F(\zeta) 
\nonumber\\&
- {9 \over 8} F(z) e^{-z/\omega} \int_{-\omega \infty}^{z} {d\zeta \over \omega} e^{\zeta/\omega} F(\zeta)  
  + {7 \over 8} e^{z/\omega} \int_{z}^{z_\Lambda} {d\zeta \over \omega} e^{-\zeta/\omega} F^2(\zeta) 
\nonumber\\&
  - (x \rightarrow \sigma; z \rightarrow z_\sigma)
  - 7 \int_{z}^{z_\sigma} {d\zeta \over \omega} F^2(\zeta) \Big] .
  \label{gluoneq}
\end{align}
\par
In order to find the solution, we substitute the Ansatz (\ref{ansatz}) into the above equations (\ref{ghosteq}) and (\ref{gluoneq}),  then perform explicitly the integration over $\zeta$, and finally comparing both sides of the resulting equations to match the coefficients and the powers on both sides.
\par
In the process of integration, 
we use the integration formula \cite{GR00}
\begin{align}
 \int^z d\zeta e^{a\zeta} \zeta^{b} = \begin{cases}
 (-1)^{-b} a^{-(1+b)} \Gamma[1+b, -az] & \text{($a \not=0$)} \\ 
 (1+b)^{-1} z^{1+b} & \text{($a=0$)} 
 \end{cases} ,
\end{align}
where $\Gamma[c,x]$ is the incomplete gamma function with the asymptotic expansion for large $|x|$:
$
 \Gamma[c,x] = x^{c-1} e^{-x} \left[ 1 + \sum_{n=1}^{\infty} {(c-1)(c-2) \cdots (c-n) \over x^n} \right] .
$
In particular, for large $z$ and $a\not=0$, we can use the formula,
\begin{align}
 \int^z d\zeta e^{a\zeta} \zeta^{b} =  
  a^{-1}  e^{az} z^{b} \left[ 1 + \sum_{\ell=1}^{\infty} 
(az)^{-\ell} 
\prod_{j=0}^{\ell-1} (j-b) \right] . 
\end{align}
In order to use the UV or IR asymptotic solutions in the integrand, we perform the decomposition: 
$
 \int_{0}^{x}dy f(y) =  \int_{\Lambda^2}^{x}dy f(y) + \int_{0}^{\Lambda^2}dy f(y)
$
for the UV limit $x \rightarrow \Lambda^2 (\rightarrow \infty)$,  
or
$\int_{x}^{\Lambda^2}dy f(y) = \int_{x}^{\epsilon}dy f(y) + \int_{\epsilon}^{\Lambda^2}dy f(y)$
 for the IR limit $x \rightarrow \epsilon$ ($\epsilon$ is introduced to avoid IR divergence). 
The asymptotic solution can not determine the constant part,
$\int_{0}^{\Lambda^2}dy f(y)$, but it is a $p$-independent constant and does not affect the exponent.  

\par
The first observation is that {\it the Ansatz for asymptotic UV solution without power corrections ($a_n^{(\ell)}=0$ and $b_n^{(\ell)}=0$) can not satisfy the coupled SD equation}, even if we include the logarithmic corrections $c_n, d_n\not=0$.    
This fact is made sure by explicit calculations.  
This is because  the $1/(p^2)^{\ell}$ ($\ell=1,2,3$) power corrections for $x :=p^2 \rightarrow \Lambda^2 \gg 1$ are generated from the term with  whole range of integration $[0,\Lambda^2]$ which appears after  separating the integral for large $x \rightarrow \Lambda^2(\rightarrow \infty)$ as 
\begin{equation}
 {F(x) \over x^{\ell}} \int_{0}^{x} dy y^{\ell-1} F(y) 
=  {F(x) \over x^{\ell}} \int_{\Lambda^2}^{x}dy y^{\ell-1} F(y) + {F(x) \over x^{\ell}} \int_{0}^{\Lambda^2}dy y^{\ell-1} F(y) ,
\end{equation}
in the RHS of the gluon equation $F^{-1}(x)$. 
Due to the term 
$- {9F(x) \over 8x} \int_{0}^{\Lambda^2}dy F(y)$, a solution without a leading power correction $O(1/p^2)$ is not allowed in the truncated SD equation in question.%
\footnote{It is pedagogic to see whether it may happen that the integral $\int_{0}^{x} dy y^{\ell-1} F(y)$ cancel the power dependence $x^{\ell}$, i.e.,  
$\int_{0}^{x} dy y^{\ell-1} F(y)=x^{\ell} f(z)$
where $f(z)$ is a polynomial of $z$. 
For this equation to be satisfied, $F(x)$ must be of the form: 
$
 F(x) = \ell f(z) + \omega f'(z) ,
$
as obtained by differentiating both sides with respect to $x$. 
Therefore, if the solution $F(x)$ is written as a polynomial of $z$ in the whole range $[0,x]$ ($x \rightarrow \infty$), the power dependence happens to cancel.  This is impossible, since we know that the IR asymptotic solution does not have such a form, see e.g. \cite{Kondo02}.  
}
  Thus, the gluon equation involves at least $1/p^2, 1/p^4, 1/p^6$ power corrections, while the ghost equation involves at least $1/p^4$ ones.  
The leading power correction originates from the gluon loop in the gluon equation.  However, the leading power correction is also generated by other terms after integration, once the Ansatz contains the power correction part.  Therefore, the actual procedure of matching the relevant terms is more complicated.%
\footnote{
In the fermion SD equation of QED and QCD, the similar situation arises if one tries to find the asymptotic solution, although the fermion SD equation can be linearized and be solved in the whole momentum region. 
However, we have checked that
the asymptotic solution agrees with the correct solution 
$\Sigma(x)=Ce^{-z}z^{-1+a} \sum_{n=0}^{N}R_n z^{-n}$  
with $z:=\ln (x/\Lambda_{QCD}^2)$,
obtained  by converting the  fermion SD equation  to the linear ordinary differential equation with appropriate boundary conditions \cite{Kondo91}.
}
Keeping these observations in mind, we proceed the calculations. 
\par
In the ghost equation, for the adopted Ansatz to be a solution, the following  relations must be satisfied (see Appendix). 
\begin{subequations}
\begin{align}
  -\delta &=   1+\gamma+\delta , 
  \label{ghost1}
\\
  1&= {3N_c \over 4}\lambda \tilde{Z}_1 AB^2 {1 \over \omega(1+\gamma+\delta)} ,
  \label{ghost2}
\\
  \delta_1 &= \gamma_1-1,
  \label{ghost3}
\\
 b_0^{(1)} &= {9N_c \over 8}\lambda \tilde{Z}_1 AB^2 a_0^{(1)} , 
\label{ghost4}
\end{align}
\label{ghostrel}
\end{subequations}
where (\ref{ghost1}) and (\ref{ghost2}) for the leading logarithm come from the matching between the LHS and the last term of RHS, i.e.,
$\int_{z}^{z_\sigma} {d\zeta \over \omega} F(\zeta) G(\zeta)$,
while (\ref{ghost3}) and (\ref{ghost4}) for the power correction are due to all the terms in the RHS.%
\footnote{
The relations (\ref{ghost3}) and (\ref{ghost4}) are obtained as follows. The matching condition between LHS and RHS implies 
$-\delta+\delta_1=\text{max}\{\gamma+\delta+\delta_1,\gamma+\delta+\gamma_1 \}$. The former choice $-\delta+\delta_1=\gamma+\delta+\delta_1$ does not hold, since $\gamma+2\delta=-1$.  Hence, the latter choice (\ref{ghost3}) must be satisfied. 
Once this identification is made, (\ref{ghost4}) immediately follows.  
}
(\ref{ghost4}) gives a ratio $b_0^{(1)}/a_0^{(1)}$. 

\par
Similarly, the gluon equation implies that the parameters must satisfy the following relations.
\begin{subequations}
\begin{align}
  -\gamma &=  1+2\delta  
  = 1+ 2\gamma  ,
  \label{gluon1}
\\ 
 1 &= {N_c \over 3}\lambda \tilde{Z}_1 AB^2 {-1 \over 2\omega(1+2\delta)}
 + {N_c \over 3}\lambda  {Z}_1 A^3 {7 \over \omega(1+2\gamma)} ,
 \label{gluon2}
\\ 
  -\gamma+\gamma_1 &=  1+2\gamma+\gamma_1  ,
  \label{gluon3}
\\ 
 1 &= {N_c \over 3}\lambda  {Z}_1 A^3 {9/8 \over \omega(1+\gamma+\gamma_1)}   .
  \label{gluon4}
\end{align}
\label{gluonrel}
\end{subequations}
Here (\ref{gluon1}) and (\ref{gluon2}) for the leading logarithm come from the matching between the LHS and the last terms in RHS,%
\footnote{The relation  (\ref{gluon1}) is obtained from
$
  -\gamma  =  \text{max}\{1+2\delta  \ (\text{ghost loop}),
1+ 2\gamma \ (\text{gluon loop}) \} .
$
 For the choice $-\gamma=1+2\delta, \delta>\gamma$ (ghost dominance), (\ref{gluon2}) must be replaced by 
$
1  = {N_c \over 3}\lambda \tilde{Z}_1 AB^2 {-1 \over 2\omega(1+2\delta)}
$.  
For this relation to be consistent with (\ref{ghost2}), we have
$-6(1+2\delta)=(4/3)(1+\gamma+\delta)=-(4/3)\delta$ and hence 
$\delta=-9/16$.  This leads to $\gamma=1/8$, contracting with $\delta>\gamma$. 
 For the choice $-\gamma=1+2\gamma, \delta<\gamma$ (gluon dominance), i.e., $\delta<\gamma=-1/3$  and (\ref{gluon2}) must be replaced by 
$
1  = {N_c \over 3}\lambda  {Z}_1 A^3 {7 \over \omega(1+2\gamma)}
$.  
This choice is excluded easily, since  
$-\gamma=1+2\delta$ from (\ref{ghost1}) leads to $\delta=\gamma$, contradicting with $\delta<\gamma$.
Once the identification (\ref{gluon1}) is made, (\ref{gluon2}) immediately follows.  
}
 i.e., 
${1 \over 2} \int_{z}^{z_\sigma} {d\zeta \over \omega}  G^2(\zeta)$
and
$- 7 \int_{z}^{z_\sigma} {d\zeta \over \omega} F^2(\zeta)$.
{\it The leading UV logarithmic behavior comes from the equal contributions of the ghost loop and the gluon loop}.  This is sharp contrast wit the IR power behavior whose dominant contribution comes from the ghost loop (ghost dominance) at least for the present truncation. 
While (\ref{gluon3}) and (\ref{gluon4}) for the power correction are due to a term 
$- {9 \over 8} e^{-z/\omega} F(z) \int_{-\omega \infty}^{z} {d\zeta \over \omega} e^{\zeta/\omega} F(\zeta) $
in the RHS.%
\footnote{
The relation (\ref{gluon3}) is obtained as follows. By comparing  both sides of the gluon equation, we obtain
$
  -\gamma+\gamma_1  = \text{max}\{  2\delta+\delta_1 \ (\text{ghost loop}), \
 1+2\gamma+\gamma_1, \gamma \ (\text{gluon loop})\}  .
$
The two conditions from the gluon loop stem from the separation of the range of integration,
$- {9 \over 8} e^{-z/\omega} F(z) \int_{-\omega \infty}^{z} {d\zeta \over \omega} e^{\zeta/\omega} F(\zeta)$,
into $[z_\Lambda,z]$ and $[-\omega \infty,z_\Lambda]$.
The matching condition is rewritten as 
$
 \gamma_1 = \text{max}\{  \gamma+2\delta+\delta_1 ,
 1+3\gamma+\gamma_1, 2\gamma \}
= \text{max}\{  \delta_1-1 ,
 \gamma_1, 2\gamma \}
= \text{max}\{  \gamma_1-2 ,
 \gamma_1, 2\gamma \}
= \text{max}\{   \gamma_1, 2\gamma \}
$
where (\ref{ghost1}), (\ref{gluon1}) and (\ref{ghost3}) were used. 
Therefore, the exponent $\gamma_1$ is determined from the gluon loop. 
When $\gamma_1>2\gamma=-2/3=-0.666\cdots$, (\ref{gluon4}) is obtained from the integral in the range $[z_\Lambda,z]$ where $1+\gamma+\gamma_1>0$.
When $\gamma_1<2\gamma$, we fall into a contradiction $\gamma_1=2\gamma$ to be excluded.
When $\gamma_1=2\gamma$, two integrals seem to give the same order of contribution.  However, this case $1+\gamma+\gamma_1=1+3\gamma=0$ leads to a new type of momentum dependence, i.e., double logarithm $\ln z \sim \ln \ln p^2$, which is beyond the Ansatz we adopted.  Therefore the case $\gamma_1=2\gamma$ is excluded in the present analysis and should be investigated elsewhere.  
Once the identification (\ref{gluon3}) is made,  we obtain
$
 a_0^{(1)} = {N_c \over 3}\lambda  {Z}_1 A^3 {9/8 \over \omega(1+\gamma+\gamma_1)}  a_0^{(1)} ,
$
which implies (\ref{gluon4}).
}
Therefore, {\it in the gluon equation, the UV power corrections originate  from the gluon loop contribution}.  We  anticipate this result, since the dimension two gluon composite operator can be inserted into the gluon loop only, see e.g., the second reference in \cite{Kondo01}. 

\par
We look for the UV asymptotic solution with leading power correction to the logarithmic solution deduced by perturbation theory.  
First, the mixed loop (\ref{ghost1}) and the ghost loop (\ref{gluon1}) yield the same relation,
\begin{equation}
 \gamma+2\delta+1=0 .
\end{equation}
  However, the gluon loop (\ref{gluon1}) implies $\gamma=-1/3$, leading to  
\begin{equation}
 \delta=-1/3=\gamma . 
\end{equation}
 This result is consistent with the relationship (\ref{gluon3}) following from the comparison between exponents of the leading power correction.
From (\ref{ghost2}), $\omega$ is obtained as
\begin{equation}
\omega=(9/4)N_c \lambda \tilde{Z}_1 AB^2 .
\end{equation}
In order to reproduce the asymptotic freedom, $\omega$ must be positive, $\omega>0$. 
 By substituting this into (\ref{gluon2}), the renormalization factor $Z_1$ is expressed as
\begin{equation}
 Z_1 = {11 \over 28} \tilde{Z}_1 {B^2 \over A^2}  . 
\end{equation}
Substitution of these relations into (\ref{gluon4}) yields the exponent
\begin{equation}
 \gamma_1 =  -\gamma -1 + 11/168   
 =  - 101/168 = - 0.60119 , 
\end{equation}
which satisfied the condition $\gamma_1>2\gamma=-2/3=-0.666\cdots$ in the footnote.
Hence, another exponent is 
\begin{equation}
 \delta_1=\gamma_1-1=-1.60119  .
\end{equation}
Finally, the ratio of two coefficients $b_0^{(1)}/a_0^{(1)}$ is obtained from (\ref{ghost4}).
It should  be remarked that the absolute values of the coefficients for the leading power corrections can not be determined from the UV asymptotic solution alone, just as the coefficient of the logarithmic correction in the IR region can not be determined from the IR asymptotic solution alone \cite{Kondo02}. 
In order to determine the coefficients,  we need to connect the UV asymptotic solution to the IR asymptotic solution and vice versa.

\par
The running coupling  constant for the gluon-ghost-antighost interaction in the UV region is given by
\begin{align}
 g^2_c(p)  
 :=&  g^2 \tilde{Z}_1 F(p^2) G^2(p) 
\nonumber\\
 =&  g^2 \tilde{Z}_1AB^2 z^{\gamma+2\delta} 
\left[ 1 +  (c_1+2d_1) z^{-1}  + (a_0^{(1)} z^{\gamma_1}+b_0^{(1)} z^{\delta_1}) e^{-z/\omega} + O\left( z^{-2}, p^{-4} \right) \right] 
\nonumber\\
 =& \left[ {(9/4)N_c \over 16\pi^2} \log {p^2 \over \sigma^2} \right]^{-1}  \left[ 1 +  (c_1+2d_1) z^{-1} + (a_0^{(1)} z^{\gamma_1}+b_0^{(1)} z^{\delta_1}) e^{-z/\omega} + O\left( z^{-2}, p^{-4} \right) \right] .
\end{align}
Another running coupling constant for the triple gluon interaction is defined by
\begin{align}
 g^2_3(p)  
 :=&  g^2  {Z}_1 F^3(p^2)   
\nonumber\\
 =&  g^2  {Z}_1 A^3 z^{3\gamma} 
\left[ 1 +  3c_1 z^{-1}  + 3a_0^{(1)} z^{\gamma_1} e^{-z/\omega} + O\left( z^{-2}, p^{-4} \right) \right] 
\nonumber\\
 =& \left[{(63/11)N_c \over 16\pi^2} \log {p^2 \over \sigma^2} \right]^{-1}  \left[ 1 +  3c_1 z^{-1} +  3a_0^{(1)} z^{\gamma_1}  e^{-z/\omega} + O\left( z^{-2}, p^{-4} \right) \right]  .
\end{align}

This solution implies the UV asymptotic freedom.
However, for the coefficient of the $\beta$ function,  it leads to $\beta_0={9 \over 4}N_c $ or $\beta_0={63 \over 11}N_c $ which are different from the perturbative result $\beta_0={11 \over 3}N_c$.  
This disagreement was resolved by taking more involved renormalization prescription in \cite{SHA97,FAR02}.  
We improve the above analysis taking into account this modification.

\section{An improvement}

Within the presented class of truncation schemes, it is impossible to satisfy both the correct one-loop scaling and the Slavnov-Taylor identity.

In order to ensure the correct one-loop scaling of the running coupling, 
Fischer, Alkofer and Reinhardt \cite{FAR02} proposed a two-parameter Ansatz, i.e., 
the gluon vertex renormalization constant $Z_1$ is replaced by the momentum dependent substitution
\begin{align}
 Z_1 \rightarrow \mathcal{Z}_1(p^2,q^2,(p-q)^2) :=
{G(q^2)^{1-a/\delta-2a} \over F(q^2)^{1+a}} 
{G((p-q)^2)^{1-b/\delta-2b} \over F((p-q)^2)^{1+b}} .
\label{newAnsatz}
\end{align}
This Ansatz is suggested using the scaling of the dressing functions extracted from the renormalization group equation \cite{SHA97}.

In the UV limit, this Ansatz yields the correct one-loop  behavior of the gluon loop, for   arbitrary $a$, $b$, as will be demonstrated below.   
The two parameters $a$ and $b$ must be chosen so that the momentum dependence of $\mathcal{Z}_1$ is as weak as possible, since $Z_1$ would be independent of momenta in a full treatment of the coupled gluon-ghost system.  

On the other hand, the IR limit strongly depends on the choice of two parameters $a, b$.  
For $b=0$, three different situation arise:  
For $a<0$, the ghost loop is dominant in the IR limit and the gluon loop is sub-leading.
For $a=0$, the gluon loop reproduce the same power as the ghost loop in the IR limit.
For $a>0$, the gluon loop becomes the leading term in the IR limit.  In this case, a solution for the coupled gluon-ghost system has not been found.   
Explicit choices taken so far are as follows. 
The choice $a=b=0$ corresponds to the truncation scheme of \cite{Bloch01}.  
Another choice 
$a=3\delta(<0), b=0$ is adopted in \cite{SHA97}.
For $b\not=0$, a choice $a=b=3\delta(<0)$ is studied in the original work 
  \cite{FAR02}.
This choice is shown to minimize the momentum dependence of $\mathcal{Z}_1$, see Appendix A of \cite{FAR02}.  

Now we adopt the Ansatz (\ref{newAnsatz}) to study the power correction in the UV region.
Repeating the similar calculations as those in the previous case, we obtain another set of relations for the power and the coefficients. 
Note that the first set of relations from the ghost equation (\ref{ghostrel}) is not affected by the substitution (\ref{newAnsatz}). 
The second set from the gluon equation (\ref{gluonrel}) is replaced by 
\begin{subequations}
\begin{align}
  -\gamma &=  1+2\delta  \ \text{(ghost loop)}
  = 1+2\delta  \ \text{(gluon loop)}, 
  \label{gluon21}
\\ 
 1 &= {N_c \over 3}\lambda \tilde{Z}_1 AB^2 {-1 \over 2\omega(1+2\delta)}
 + {N_c \over 3}\lambda  
A^{1-a-b} B^{2-a/\delta-b/\delta-2a-2b} {7 \over \omega(1+2\delta)} ,
 \label{gluon22}
\\ 
  -\gamma+\gamma_1 &= 1+2\gamma+\gamma_1  
+ \gamma(-2-a-b) + \delta (2-a/\delta-b/\delta-2a-2b) ,
  \label{gluon23}
\\ 
 1 &= {N_c \over 3}\lambda   {9/8(-1-a-b) A^{1-a-b} B^{2-a/\delta-b/\delta-2a-2b} \over \omega[1+\gamma+\gamma_1+\gamma(-2-a-b) + \delta (2-a/\delta-b/\delta-2a-2b)]}  .
  \label{gluon24}
\end{align}
\label{gluonrel2}
\end{subequations}
The relation (\ref{gluon21}) is the same as (\ref{ghost1}).
By substituting $\omega$ obtained from (\ref{ghost2}) into (\ref{gluon22}), we obtain the ratio:
$
 \gamma/\delta = -{2 \over 9} \left[ 1 - 14 \tilde{Z}_1^{-1}(AB^{2+1/\delta})^{-(a+b)} \right] .
$
By using the perturbative normalization $A=B=1$, the ratio reduces to    
$\gamma/\delta =26/9$ which implies 
\begin{equation}
 \gamma = - 13/22 =-0.590909 , \quad 
 \delta = -  9/44  = - 0.204545.
\end{equation}
Returning to (\ref{ghost2}) yields   
\begin{equation}
\omega=(11/3)N_c \lambda \tilde{Z}_1 AB^2
= (11/3)N_c \lambda .
\label{omega}
\end{equation}
The relation (\ref{gluon23}) is trivially satisfied under the relation (\ref{gluon21}) and does not give any restriction.
For $a_0^{(1)}\not=0$, the final relation (\ref{gluon24}) with the substitution (\ref{omega}) gives the exponent of power correction: 
\begin{equation}
 \gamma_1 = -\gamma-1- \gamma(-2-a-b) - \delta (2-a/\delta-b/\delta-2a-2b) - 9/88(1+a+b) ,
\end{equation}
whose value depends on the choice of $a, b$:
\begin{equation}
 \gamma_1 = -0.74845 (a=3\delta=b) ,
-0.323089 (a=3\delta, b=0) ,
 +0.102273 (a=0,b=0)  .
\end{equation}
The result (\ref{omega}) shows that the coefficient $\beta_0={11/3N_c \over 16\pi^2}$ of the $\beta$ function is correctly reproduced  by the  identification $\beta_0 g^2=\omega$.

\section{Comparison with OPE result}

The UV asymptotic solution obtained above for the gluon and ghost propagators with power corrections should be compared with those obtained by the OPE. 
To one-loop order, the renormalized gluon propagator has the OPE \cite{Boucaudetal00,Boucaudetal02b}
\begin{align}
  D_T^R(p^2) = {1 \over p^2} \left[ A_0(p)  
+ {A_2(p) \over p^2} {\langle \mathscr{A}_R(0)^2 \rangle \over 4(N_c^2-1)} \right] ,
\end{align}
where the Wilson coefficients are given by
\begin{align}
   A_0(p) =& 
\left( g^2(p)/g^2(\mu) \right)^{{\gamma_0 \over \beta_0}} 
\sim  \left( \ln {p \over \Lambda_{QCD}}/\ln {\mu \over \Lambda_{QCD}}  \right)^{-{13 \over 22}} ,
\nonumber\\
 A_2(p) =& N_c g^2(p) \left( g^2(p)/g^2(\mu) \right)^{-{\hat{\gamma}_0 \over \beta_0}} 
\sim  \left(  \ln {p \over \Lambda_{QCD}}/\ln {\mu \over \Lambda_{QCD}} \right)^{-{35 \over 44}} ,
\end{align}
with the coefficients of the $\beta$ function and anomalous dimension of the gluon field and the composite gluon field:
$
  \beta_0 = {11 \over 3}N_c, \
  \gamma_0 = {13 \over 6} N_c,  \
\gamma_{A^2} = {35 \over 12}N_c,  \
  \hat{\gamma}_0 = \gamma_{A^2}-\gamma_0 = {3 \over 4}N_c .
$
As already mentioned above, the exponent of logarithmic dependence completely agrees with the perturbative result: $\gamma=\gamma_0/\beta_0$. 
We find the exponent of logarithmic correction $-(1-{\hat{\gamma}_0 \over \beta_0})=-35/44=-0.795455$ from OPE.
The best fit is obtained  $\gamma+\gamma_1 =-0.913998$ for the choice $a=3\delta, b=0$ among the cases treated in the previous section (The naive treatment yields the value $-0.93452$.)  
Thus, the logarithmic momentum dependence of the power correction obtained from the UV asymptotic solution of the SD equation can reproduce the prediction based on the Wilson coefficient $A_2$ of OPE to one-loop order. 

On the other hand,  the renormalized ghost propagator has the OPE \cite{Kondo01} 
\begin{align}
  G_{gh}^R(p^2) = {1 \over p^2} \left[ B_0(p)  
+ {B_2(p) \over p^2} {\langle \mathscr{A}_R(0)^2 \rangle \over 4(N_c^2-1)} \right] ,
\end{align}
where 
\begin{align}
   B_0(p) =& \left( g^2(p)/g^2(\mu)  \right)^{{\tilde{\gamma}_0 \over \beta_0}} 
\sim  \left(  \ln {p \over \Lambda_{QCD}}/\ln {\mu \over \Lambda_{QCD}}  \right)^{-{9 \over 44}} ,
\nonumber\\
 B_2(p) =& N_c g^2(p) \left( g^2(p)/g^2(\mu) \right)^{-{\hat{\tilde{\gamma}}_0 \over \beta_0}} 
\sim  \left(  \ln {p \over \Lambda_{QCD}}/\ln {\mu \over \Lambda_{QCD} } \right)^{-{9 \over 22}} ,
\end{align}
with
$
  \beta_0 = {11 \over 3}N_c, \
  \tilde{\gamma}_0 = {3 \over 4} N_c,  \
\gamma_{A^2} = {35 \over 12}N_c,  \
  \hat{\tilde{\gamma}}_0 = \gamma_{A^2}-\tilde{\gamma}_0 = {13 \over 6}N_c . 
$
Hence, the solution of SD equation exhibits large deviation $\delta+\delta_1 =-1.913998$ from the OPE result $-(1-{\hat{\tilde{\gamma}}_0 \over \beta_0})= -{9 \over 22}=-0.409091$ for the ghost propagator.  
This is the issue to be resolved in the near future. 

\par
The running coupling  constant for the gluon-ghost-antighost interaction is given by
\begin{align}
 g^2_c(p)  
 =&  g^2 \tilde{Z}_1 F(p^2) G^2(p) 
\nonumber\\
 =&  g^2 \tilde{Z}_1AB^2 z^{\gamma+2\delta} 
\left[ 1 +  (c_1+2d_1) z^{-1}  + (a_0^{(1)} z^{\gamma_1}+b_0^{(1)} z^{\delta_1}) e^{-z/\omega} + O\left( z^{-2}, p^{-4} \right) \right] 
\nonumber\\
 =& g^2_{pert}(p)  \left[ 1 +  (c_1+2d_1) z^{-1} + (a_0^{(1)} z^{\gamma_1}+b_0^{(1)} z^{\delta_1}) e^{-z/\omega} + O\left( z^{-2}, p^{-4} \right) \right]  ,
\end{align}
while the running coupling constant for the triple gluon interaction is modified as
\begin{align}
 g^2_3(p)  
 =&  g^2  \mathcal{Z}_1 F^3(p^2)   
= g^2 F(p^2)^{1-a-b} G(p^2)^{2-a/\delta-b/\delta-2a-2b}
\nonumber\\
 =&  g^2 A^{1-a-b} B^{2-a/\delta-b/\delta-2a-2b} z^{\gamma(1-a-b)+\delta(2-a/\delta-b/\delta-2a-2b)} 
\nonumber\\ & \times
\left[ 1 +  3c_1 z^{-1} +  3a_0^{(1)} z^{\gamma_1}  e^{-z/\omega} + O\left( z^{-2}, p^{-4} \right) \right] 
\nonumber\\
 =& g^2_{pert}(p)  \left[ 1 +  3c_1 z^{-1} +  3a_0^{(1)} z^{\gamma_1}  e^{-z/\omega} + O\left( z^{-2}, p^{-4} \right) \right]  ,
\end{align}
where
$
g^2_{pert}(p) 
 := [{(11/3)N_c \over 16\pi^2} \log {p^2 \over \Lambda_{QCD}^2}]^{-1} .
$
Two running coupling constants do not agree to each other beyond the one-loop expression, since the Slavnov-Taylor identity is sacrificed by the bare vertex approximation. 
\par
The OPE for the coupling constant of triple gluon interaction \cite{Boucaudetal00} shows 
\begin{align}
  g^2(p) = g^2_{pert}(p) \left\{ 1 + R \left( \ln (p/\Lambda_{QCD}) \right)^{(\gamma_0+\hat{\gamma}_0)/\beta_0-1} \langle \mathscr{A}_\mu \mathscr{A}_\mu  \rangle/p^2   \right\} ,
\end{align}
where $(\gamma_0+\hat{\gamma}_0)/\beta_0-1=-9/44=-0.204545$
and
$
  R :=   {18\pi^2 \over \beta_0 (N_c^2-1)}  \left( \ln (\mu/\Lambda_{QCD}) \right)^{-(\gamma_0+\hat{\gamma}_0)/\beta_0} .
$
For the running coupling, therefore, the agreement for the exponent $\gamma_1, \delta_1$ between the SD equation with the OPE is not so good.  This is because in our truncation of the SD equation the bare vertex approximation is adopted, while in the OPE the vertex correction was taken into account to the leading logarithm.  
 
\section{Conclusion and discussion}

We have proposed a new Ansatz for obtaining the asymptotic solution of the coupled SD equation for the gluon and ghost form factors (or propagators) in Yang-Mills theory.  
We have found  a set of consistent UV asymptotic solutions with the leading power corrections $O(1/p^2)$ for the gluon and ghost form factors: 
$
  F(z) = A z^{\gamma} \sum_{n=0}^{N} c_n z^{-n} 
   +  e^{- z/\omega} z^{\gamma+\gamma_1}   a_0^{(1)}  ,
$
$
  G(z) = B z^{\delta} \sum_{n=0}^{N} d_n z^{-n} 
   +  e^{- z/\omega} z^{\delta+\delta_1}   b_0^{(1)}  ,
$
within the framework of the truncated SD equation of Yang-Mills theory in the Landau gauge.

\par
Within the present truncation of the SD equation, we have calculated the exponents of logarithmic dependence in the power correction term $O(1/p^2)$.
The momentum dependence of the power correction part in the gluon propagator agrees well with the OPE result by taking into account a suitable modification for the triple gluon vertex, while the ghost propagator does not show good agreement.  
In this paper we have adopted the bare vertex function and the simple modification for truncating the SD equation.  Therefore, it will be important to see whether our qualitative result remains true even after  the vertex function is improved so as to satisfy the Slavnov-Taylor identity.  This might give a clue to resolve the discrepancy between the OPE and the SD equation especially for the ghost propagator. 
\par
Thus, the SD equation suggests the existence of the leading power correction to the UV asymptotic solution.  The existence of the leading power correction is consistent with the existence of vacuum condensate with mass dimension 2, i.e., $\langle \mathscr{A}_\mu^2 \rangle \not=0$, 
as argued recently by several groups \cite{Boucaudetal00,GSZ01,Kondo01}.  
It turns out that the existence of power corrections with additional logarithmic dependence of momenta leads to the existence of sub-leading terms with logarithmic dependence.  
\par
A way to determine the numerical value of non-vanishing coefficient $a_0^{(1)}$ or $b_0^{(1)}$ and thereby the numerical value of the condensation $\langle \mathscr{A}_\mu^2 \rangle$
is to perform the numerical calculation which enables us to connect the UV solution into the IR one. 
It will clarify how the UV solution with the vacuum condensate 
$\langle \mathscr{A}_\mu^2 \rangle$ is  connected to the IR solution 
in which the new mass scale $M_G$, e.g., the Gribov mass (scale)  is expected to appear. 
 Another way is to require the (Borel) summability of the infinite series $N=\infty$ where the consistency of the theory in presence of IR renormalon can be used to estimate the numerical value of the vacuum condensate $\langle \mathscr{A}_\mu^2 \rangle$. 
It is important to notice that the mixing between the leading power correction and the logarithmic parts does not occur.     
These issues will be studied in a subsequent paper. 
\par
It is in principle possible to extend this work to obtain the sub-leading power correction to the order $O(1/p^4)$, corresponding to the vacuum condensate of mass dimension 4, e.g., 
$\langle \mathscr{F}_{\mu\nu}^2 \rangle \not=0$. 
Finally, it will be interesting to study the UV asymptotic solution in other covariant gauges, e.g., the Maximal Abelian (MA) gauge \cite{tHooft81} where a different type of vacuum condensate of mass dimension 2 is expected to occur, i.e., simultaneous Bose-Einstein condensation of off-diagonal gluon and ghost \cite{Schaden99,Kondo01}.

\appendix
\section{Integration of SD equation}

After performing the momentum integration, the ghost equation reads
\begin{align}
 & Bz^{\delta} G^{-1}(z)
= \sum_{N=0}^{M} G_N z^{-N} 
 + e^{-z/\omega} z^{\delta_1} \sum_{N=0}^{M} Q_N^{(1)} z^{-N}
 + \cdots 
\nonumber\\
=& -{3N_c \over 4} \nu \sum_{n,m=0}^{M} \Big\{
c_n d_m \frac{1}{2} z^{\gamma+2\delta-n-m} 
\sum_{\ell=0}^{M} z^{-\ell} \left(\frac{\omega}{2}\right)^{\ell} \prod_{i=0}^{\ell-1}(m-\delta+i) 
\nonumber\\ & \quad\quad\quad\quad\quad\quad
- c_n d_m \frac{1}{\omega(1+\gamma+\delta-n-m)} z^{1+\gamma+2\delta-n-m}
\nonumber\\ & \quad\quad\quad\quad\quad\quad
+ e^{-z/\omega} \Big[
 c_n d_m b_m^{(1)} z^{\gamma+2\delta+\delta_1-n-m}
\sum_{\ell=0}^{M} z^{-\ell} \omega^{\ell} \prod_{i=0}^{\ell-1}(m-\delta-\delta_1+i) 
\nonumber\\ & \quad\quad\quad\quad\quad\quad\quad\quad\quad
+  c_n d_m a_n^{(1)}  z^{\gamma+2\delta+\gamma_1-n-m} \frac{1}{2}
\sum_{\ell=0}^{M} z^{-\ell} \left(\frac{\omega}{2}\right)^{\ell} \prod_{i=0}^{\ell-1}(m-\delta+i) 
\nonumber\\ & \quad\quad\quad\quad\quad\quad\quad\quad\quad
+  c_n d_m a_n^{(1)}  z^{\gamma+2\delta+\gamma_1-n-m}
\sum_{\ell=0}^{M} z^{-\ell} (-\omega)^{\ell} \prod_{i=0}^{\ell-1}(n+m-\gamma-\delta-\gamma_1+i) 
\nonumber\\ & \quad\quad\quad\quad\quad\quad\quad\quad\quad
+  c_n d_m b_m^{(1)}  z^{\gamma+2\delta+\delta_1-n-m}
\sum_{\ell=0}^{M} z^{-\ell} (-\omega)^{\ell} \prod_{i=0}^{\ell-1}(n+m-\gamma-\delta-\delta_1+i) 
\Big] 
\Big\}
,
\end{align}
where we have defined
$
  \nu := \lambda \tilde{Z}_1 AB^2 
$.
The gluon equation reads
\begin{subequations}
\begin{align}
 & Az^{\gamma} F^{-1}(z)
= \sum_{N=0}^{M} F_N z^{-N} 
 + e^{-z/\omega} z^{\gamma_1} \sum_{N=0}^{M} P_N^{(1)} z^{-N}
 + \cdots 
\nonumber\\
=& {N_c \over 3} \nu \sum_{n,m=0}^{M} \Big\{
 z^{\gamma+2\delta-n-m} d_n d_m  \left[ -\frac{1}{3}
\sum_{\ell=0}^{M} z^{-\ell} \left(\frac{\omega}{3}\right)^{\ell} \prod_{i=0}^{\ell-1}(m-\delta+i) 
+ \frac{3}{4}
\sum_{\ell=0}^{M} z^{-\ell} \left(\frac{\omega}{2}\right)^{\ell} \prod_{i=0}^{\ell-1}(m-\delta+i) \right] 
\nonumber\\ & \quad\quad\quad\quad
- z^{1+\gamma+2\delta-n-m} d_n d_m \frac{1}{2} \frac{1}{\omega(1+2\delta-n-m)} 
\nonumber\\ & \quad\quad\quad\quad
+ e^{-z/\omega} z^{\gamma+2\delta+\delta_1-n-m}   \Big[
-  d_n d_m b_m^{(1)} \frac{1}{2} 
\sum_{\ell=0}^{M} z^{-\ell} \left(\frac{\omega}{2}\right)^{\ell} \prod_{i=0}^{\ell-1}(m-\delta-\delta_1+i) 
\nonumber\\ & \quad\quad\quad\quad\quad\quad\quad\quad\quad\quad 
-   d_n d_m b_n^{(1)} \frac{1}{3}  
\sum_{\ell=0}^{M} z^{-\ell} \left(\frac{\omega}{3}\right)^{\ell} \prod_{i=0}^{\ell-1}(m-\delta+i) 
\nonumber\\ & \quad\quad\quad\quad\quad\quad\quad\quad\quad\quad 
+  d_n d_m b_m^{(1)} \frac{3}{2}  
\sum_{\ell=0}^{M} z^{-\ell} \omega^{\ell} \prod_{i=0}^{\ell-1}(m-\delta-\delta_1+i) 
\nonumber\\ & \quad\quad\quad\quad\quad\quad\quad\quad\quad\quad 
+  d_n d_m b_n^{(1)} \frac{3}{4}  
\sum_{\ell=0}^{M} z^{-\ell} \left(\frac{\omega}{2}\right)^{\ell} \prod_{i=0}^{\ell-1}(m-\delta+i) 
\nonumber\\ & \quad\quad\quad\quad\quad\quad\quad\quad\quad\quad 
+  d_n d_m (b_n^{(1)}+b_m^{(1)}) \frac{1}{2}  
\sum_{\ell=0}^{M} z^{-\ell} (-\omega)^{\ell} \prod_{i=0}^{\ell-1}(n+m-2\delta-\delta_1+i) 
\Big] 
\Big\}
\end{align}
\begin{align}
&\nonumber\\
+& {N_c \over 3} \nu{}' \sum_{n,m=0}^{M} \Big\{
 z^{3\gamma-n-m} c_n c_m  \Big[ 
 \frac{7}{6}
\sum_{\ell=0}^{M} z^{-\ell} \left(\frac{\omega}{3}\right)^{\ell} \prod_{i=0}^{\ell-1}(m-\gamma+i) 
- \frac{17}{4} 
\sum_{\ell=0}^{M} z^{-\ell} \left(\frac{\omega}{2}\right)^{\ell} \prod_{i=0}^{\ell-1}(m-\gamma+i) 
\nonumber\\ & \quad\quad\quad\quad\quad\quad
- \frac{9}{8} 
\sum_{\ell=0}^{M} z^{-\ell}  \omega^{\ell} \prod_{i=0}^{\ell-1}(m-\gamma+i) 
+ \frac{7}{8} 
\sum_{\ell=0}^{M} z^{-\ell} (-\omega)^{\ell} \prod_{i=0}^{\ell-1}(m+n-2\gamma+i) 
 \Big]
\nonumber\\ & \quad\quad\quad\quad 
+ z^{1+3\gamma-n-m} c_n c_m   \frac{7}{\omega(1+2\gamma-n-m)} 
\nonumber\\ & \quad\quad\quad\quad 
+ e^{-z/\omega} z^{3\gamma+\gamma_1-n-m} \Big[
+  c_n c_m  a_m^{(1)} \frac{7}{4} 
\sum_{\ell=0}^{M} z^{-\ell} \left(\frac{\omega}{2}\right)^{\ell} \prod_{i=0}^{\ell-1}(m-\gamma-\gamma_1+i) 
\nonumber\\ & \quad\quad\quad\quad\quad\quad\quad\quad\quad\quad 
+   c_n c_m  a_n^{(1)} \frac{7}{6} 
\sum_{\ell=0}^{M} z^{-\ell} \left(\frac{\omega}{3}\right)^{\ell} \prod_{i=0}^{\ell-1}(m-\gamma+i) 
\nonumber\\ & \quad\quad\quad\quad\quad\quad\quad\quad\quad\quad 
-   c_n c_m  a_m^{(1)} \frac{17}{2}  
\sum_{\ell=0}^{M} z^{-\ell} \omega^{\ell} \prod_{i=0}^{\ell-1}(m-\gamma-\gamma_1+i) 
\nonumber\\ & \quad\quad\quad\quad\quad\quad\quad\quad\quad\quad 
-   c_n c_m  a_n^{(1)} \frac{17}{4}  
\sum_{\ell=0}^{M} z^{-\ell} \left(\frac{\omega}{2}\right)^{\ell} \prod_{i=0}^{\ell-1}(m-\gamma+i) 
\nonumber\\ & \quad\quad\quad\quad\quad\quad\quad\quad\quad\quad 
-   c_n c_m  a_n^{(1)} \frac{9}{8}  
\sum_{\ell=0}^{M} z^{-\ell} \omega^{\ell} \prod_{i=0}^{\ell-1}(m-\gamma+i) 
\nonumber\\ & \quad\quad\quad\quad\quad\quad\quad\quad\quad\quad 
+   c_n c_m  (a_n^{(1)}+a_m^{(1)}) \frac{7}{16} 
\sum_{\ell=0}^{M} z^{-\ell} \left(-\frac{\omega}{2}\right)^{\ell} \prod_{i=0}^{\ell-1}(n+m-2\gamma-\gamma_1+i) 
\nonumber\\ & \quad\quad\quad\quad\quad\quad\quad\quad\quad\quad 
-   c_n c_m  (a_n^{(1)}+a_m^{(1)}) 7
\sum_{\ell=0}^{M} z^{-\ell} (-\omega)^{\ell} \prod_{i=0}^{\ell-1}(n+m-2\gamma-\gamma_1+i) 
\Big] 
\nonumber\\ & \quad\quad\quad\quad 
 -\frac{9}{8} e^{-z/\omega} z^{1+3\gamma+\gamma_1-n-m} c_n c_m   a_m^{(1)}  
\frac{1}{\omega(1+\gamma+\gamma_1-m)}
\Big\}
\nonumber\\ & 
- {N_c \over 3} \nu{}' \sum_{n=0}^{M} e^{-z/\omega}  z^{2\gamma-n} c_n \frac{9}{8} \int_{-\omega\infty}^{z_\Lambda} {d\zeta \over \omega} e^{\zeta/\omega} F(\zeta)/A 
,
\end{align}
\end{subequations}
where we have defined
$ 
  \nu' := \lambda Z_1 A^3 = {A^2 \over B^2}{Z_1 \over \tilde{Z}_1} \nu ={11 \over 28}\nu  . 
$ 
The above integrations were performed under the assumption that
$1+\gamma+\delta,1+2\gamma,1+2\delta,1+\gamma+\gamma_1$
are not integers. 
 In the LHS, the coefficients are obtained as
\begin{align}
 G_0 =& 1, \quad 
 G_1 = -d_1, \quad 
 G_2 = -d_2+d_1^2, \quad 
 G_3 = -d_3+2d_1 d_2-d_1^3 , \quad 
\nonumber\\   
 G_4 =& -d_4+2d_1 d_3-3d_1^2 d_2+d_1^4 , 
  \cdots , 
\nonumber\\
 F_0 =& 1, \quad 
 F_1 =  -c_1,  \quad 
 F_2 = -c_2+c_1^2 , \quad 
 F_3 = -c_3+2c_1 c_2-c_1^3 , \quad 
\nonumber\\    
 F_4 =& -c_4+2c_1 c_3-3c_1^2 c_2+c_1^4 , 
 \cdots ,
\nonumber\\    
 Q_0^{(1)} =& - b_0^{(1)} , \quad
Q_0^{(1)} = d_1(2b_0^{(1)}-b_1^{(1)}), \quad
 P_0^{(1)} =  - a_0^{(1)} , \quad 
P_1^{(1)} = c_1(2a_0^{(1)}-a_1^{(1)}), 
 \cdots .
\end{align}

\section*{Acknowledgments}
The author would like to thank Olivier Pene for kind hospitality in LPT, Universit\'e de Paris XI and 
Reinhardt Alkofer, Jacques C.R. Bloch, Kurt Langfeld and Hugo Reinhardt for kind hospitality in T\"ubingen University. 
He is grateful to Takahito Imai for drawing the figure and Takeharu Murakami for pointing out errors in the calculation. 
This work is supported by Sumitomo Foundations and by 
Grant-in-Aid for Scientific Research from the Ministry of
Education, Science and Culture: (B)13135203 and (C)14540243.


\end{document}